\documentclass[a4paper,12pt]{article}
\usepackage{authblk}

\usepackage[numbers]{natbib}
\usepackage{amsmath}
\usepackage{endnotes}
\usepackage{color}
\usepackage{enumerate}

\usepackage{longtable, lscape} 
\usepackage{graphicx} 
\usepackage[top=2.5cm,bottom=2.5cm,left=2cm,right=2cm]{geometry} 

\usepackage[breaklinks=true]{hyperref} 

\usepackage[version=3]{mhchem} 

\usepackage{color}
\usepackage{longtable, lscape} 
\usepackage{comment}
\author{Andrey I. Frolov}
\affil{Institute of Solution Chemistry, Russian Academy of Sciences, Akademicheskaya St. 1,
153045 Ivanovo, Russia. E-mail: andrey.i.frolov@mail.ru
\\
\textit{Current address:} Sanofi R\&D, 13 quai Jules Guesde, 94403 Vitry-sur-Seine, France}

\title{\textbf{Theory of Solutions in Energy Representation in NPT-ensemble: Derivation Details}}

\begin{document}

\maketitle

\begin{abstract}
Theory of solutions in energy representation (ER method) developed by Matubayasi and
Nakahara provides with an approximate way of calculating solvation free energies (or, identically,
the excess chemical potentials) from atomistic simulations. In this document we provide some
derivation details of this, to our opinion, theoretically involved method, which will help a
non-specialist to follow. There are three points which differ this document from a regular
textbook on statistical mechanics or research articles:

\begin{enumerate}

\item Derivation is detailed and all approximations are explicitly stated; 
\item Statistical mechanics derivations are performed in NPT-ensemble;  
\item We perform the derivations for the case when a molecule is represented as a set of
(atomic) sites interacting via spherically symmetric potentials (a classical Force Field
representation).

\end{enumerate}

In ER method, a new collective coordinate is introduced - the interaction
energy of a solute and a solvent molecule. The excess chemical potential is expressed as a
functional of the solute-solvent density distribution defined over the collective variable. The
functional can be approximated by the Percus's method of functional expansion, which leads to the
end-point (not dependent on the $\lambda$-coupling path) free energy expression. 

As a side result, we prove that the solvation free energy is always equivalent to the excess (over
ideal) chemical potential, and not only at infinite dilution or when internal molecule degrees of
freedom are not affected by solvation as it is sometimes wrongly believed. 

\end{abstract}

\newpage
\tableofcontents

\section{Introduction}

We provide detailed derivation of the theory of solutions in energy representation
(ER method) developed by Matubayasi and
Nakahara \cite{Matubayasi2000,Matubayasi2002,Matubayasi2003,Matubayasi2006,Matubayasi????}. ER
method provides with an approximate way of calculating solvation free energies (or, identically,
the excess chemical potentials) from atomistic simulations. The method can be seen as bridge between
the molecular simulations and the classical density
functional theory (DFT). It is quite common nowadays to model molecular interaction on the level of
the classical force field approximation, which implies that a molecule is represented as a set
of (atomic) sites interacting via spherically symmetric potentials. 

In the first part of the manuscript we
define and derive the expression for solvation free energy (SFE) in NPT-ensemble for the case of
classical force field representation of molecular interactions. We prove that SFE is
identical to the excess chemical potential. Also, we obtain the Kirkwood's charging formula
expressing
the excess chemical potential via the solute-solvent density distribution.

In the second part of the manuscript we provide details on some important relations of ER method for
the case of NPT-ensemble. In ER method, a new collective coordinate is introduced - the interaction
energy of a solute and a solvent molecule. We show that the Kirkwood's charging formula is valid
also for the solute-solvent distribution function in energy representation. Later this expression
is reformulated as a functional of density distribution. The Percus’s method of functional
expansion is used to obtain the hypernetted chain (HNC)- and Percus-Yevick (PY) -like approximations
of this functional. The final
formula for the excess chemical potential heuristically combines expressions from
different approximations and employs different input functions.

\section{Excess chemical potential in NPT-ensemble}

\subsection{Some definitions}

We consider a system with $N_s=1$ solute and $N_w$ solvent molecules in
isothermo-isobaric ensemble (NPT-ensemble).

We describe the interactions
between the molecules in the force field approximation at the level of classical mechanics. Each
molecule is represented as a set of atoms (better to say interaction sites), which interact with
each other via bonded and nonbonded potentials present in the given force field (e.g. OPLS,
CHARMM, AMBER, etc). Each
interaction site is considered as a separate object, which has its own translational
degrees of freedom and translational partition function. Therefore, when we talk about a set of 
coordinates which define the position of a molecule $\mathbf{x}_i$ we mean the positions of all
atoms which belong to this molecule, where index $\alpha$ runs over all atoms ($n_\text{t}$) of the
molecule of type $\text{t}$:

\begin{equation}
\mathbf{x}_i = \{ \mathbf{r}_{i,\alpha} \}_{\alpha=1}^{n_\text{t}}
\label{eq:bx}
\end{equation}
where $\text{t}$ is the molecule type, e.g. $s$ denotes solute, $w$ denotes solvent, and the
coordinates of
atom $\alpha$ of i$^{\text{th}}$ molecule:
$$
\mathbf{r}_{i,\alpha} = \{ x_{i,\alpha},y_{i,\alpha},z_{i,\alpha} \}
$$

Each atom has its own momentum: $\mathbf{p}_{i,\alpha}$

\subsection{Parametrized Hamiltonian}

Here and after we mostly adopt the notations used in the Appendix of the Shirts et al.
publication \cite{Shirts2003}.

The excess chemical potential can be calculated in the process of gradual switching on the
intermolecular interactions between a solute molecule and the solvent. We introduce $\lambda$
parameter which controls the degree of coupling between the solute and solvent molecules, such that
when $\lambda$=0 the interactions are absent and when $\lambda$=1 interactions are at full
coupling. Since, only solute-solvent interaction potential $u_{sw,\lambda}
(\mathbf{x}_s,\mathbf{x}_{w,i})$ depends on $\lambda$, the potential energy function of the system
can be written as follows:

\begin{equation}
U_\lambda (\mathbf{x}_s,\mathbf{x}_w^{N_w}) 
= 
\Psi (\mathbf{x}_s) 
+
\sum_{i=1}^{N_w} u_{sw,\lambda}
(\mathbf{x}_s,\mathbf{x}_{w,i}) 
+
U_{ww}(\mathbf{x}^{N_w})
\label{eq:Ul}
\end{equation}
where subscript $s$ denotes solute, subscript $w$ denotes solvent, $\Psi (\mathbf{x}_s)$ is the
potential energy of the solute molecule, $\mathbf{x}_{w,i}$ is the position of $i^{\text{th}}$
solvent molecule, $N_w$ is the number of solvent molecules, $u_{sw,\lambda}$ is the
$\lambda$-dependent solute-solvent interaction potential, $U_{ww}$ is the potential energy of the
solvent molecules, $\mathbf{x}^{N_w}$ is the short notation of positions of all solvent 
molecules. 

The total Hamiltonian can be written as:
\begin{equation}
H_\lambda = K(\mathbf{p}_s,\mathbf{p}_w^{N_w}) + U_\lambda
(\mathbf{x}_s,\mathbf{x}_w^{N_w})
\label{eq:Hl}
\end{equation}

where the kinetic energy is written as:
\begin{equation}
K(\mathbf{p}_s,\mathbf{p}_w^{N_w}) = \sum_{\alpha=1}^{n_s} \frac{p_{s,\alpha}^2}{2 m_{s,\alpha}} + 
 \sum_{i=1}^{N_w} \sum_{\alpha=1}^{n_w} \frac{p_{w,i,\alpha}^2}{2 m_{w,\alpha}}
\label{eq:K}
\end{equation}
where $m_{s,\alpha}$ and $m_{w,\alpha}$ are the masses of $\alpha^{\text{th}}$ atoms of solute and
solvent molecules, correspondingly.

\subsection{Partition functions with non-parameterized Hamiltonian}

Keeping in mind that we consider the system with a single solute molecule $N_s=1$, we will write
explicitly the terms with $N_s$ in the derivations. Later this will help us to show that the SFE
is always equal to the excess chemical potential.

\subsubsection{Case of solution}

The partition function in NPT ensemble can be written as:
\begin{equation}
\Delta (N_s,N_w,P,T) = \int_0^\infty d \left( \frac{V}{V^\prime} \right) e^{-\beta PV}
Q(N_s,N_w,V,T)
\label{eq:Dnpt}
\end{equation}
where $V^{\prime}$ is an arbitrary constant which makes the partition function dimensionless, 
$\beta$ is $(k_BT)^{-1}$, $k_B$ is the Boltzmann constant, $Q(N_s,N_w,V,T)$ is the canonical
partition function, which has the following form:

$$
Q(N_s,N_w,V,T) = 
\frac
{1}
{h^{3 N_w n_w} N_w! h^{3 N_s n_s} N_s! }
\int_{-\infty}^{+\infty} d\mathbf{p}_s^{N_s} d\mathbf{p}_w^{N_w} \int_V d\mathbf{x}_s^{N_s}
d\mathbf{x}_w^{N_w}
  \exp \left[ -\beta H
(\mathbf{p}_s^{N_s},\mathbf{p}_w^{N_w},\mathbf{x}_s^{N_s},\mathbf{x}_w^{N_w}) 
\right]
$$
where $h$ is the Planck's constant. Multiplication by $h^{-1}$ serves as a quantum correction
for purely classical partition function \cite{Ben-Naim2006}. The factorials of number of
atoms in the system appear due to indistinguishably of atoms belonging to the molecules of the
same type. Each integration symbol denotes integration over multiple coordinates. Differential
$d\mathbf{x}_w^{N_w}$ is the short notation for $d\mathbf{x}_{w,1}...d\mathbf{x}_{w,N_w}$. Symbol
$V$ at the integration sign - $\int_V$ - reflects that integration limits are bound by the system's
volume. 


In the case of classical statistical mechanics the momenta degrees of freedom are independent and
can be analytically integrated \cite{Hansen1991}:
$$
 h^{-3}  \int_{-\infty}^{+\infty} d\mathbf{p}_{x,\alpha} e^{-\beta ( \frac{p_{x,\alpha}^2}{2
m_{x,\alpha}} ) } =
\left[ \frac{h^2}{2 \pi m_{x,\alpha} k_BT} \right]^{-1.5} = \Lambda_{x,\alpha}^{-3}
$$
where $\Lambda_{x,\alpha}$ is the thermal de Broglie wavelength for atom $\alpha$ in molecule of
type $x$.

Therefore we get:
\begin{equation}
Q(N_s,N_w,V,T) =
 \prod_{\alpha=1}^{n_s} \frac{\Lambda_{s,\alpha}^{-3N_s}}{N_s!}  
 \prod_{\alpha=1}^{n_w} \frac{\Lambda_{w,\alpha}^{-3N_w}}{N_w!} \cdot  
Z (N_s,N_w,V,T)
\label{eq:Q}
\end{equation}

where $Z$ is the configuration integral of the system:
\begin{equation}
Z(N_s,N_w,V,T) = \int_V d\mathbf{x}_s^{N_s} d\mathbf{x}_s^{N_w}   
\exp 
\left[ 
    -\beta 
    U (\mathbf{x}_s^{N_s},\mathbf{x}_w^{N_w})
\right]
\label{eq:Z}
\end{equation}


The Gibbs free energy is:
$$
G(N_s,N_w,P,T) = -k_BT \ln \Delta (N_s,N_w,P,T)
$$

The chemical potential of solute in the system can be written as:
\begin{equation}
\mu = G(N_s,N_w,P,T) - G(N_s-1,N_w,P,T) 
= -k_BT \ln 
\frac{\Delta (N_s,N_w,P,T)}
     {\Delta (N_s-1,N_w,P,T)}
\label{eq:mu}
\end{equation}

\subsubsection{Case of ideal gas}

For later derivations we will use the expression for the chemical potential of
non-interacting solute molecules at given $T$, $V$ and $N_s$. Therefore, we derive it here. 
Firstly, let us find the configuration integral for a single solute molecule:
\begin{equation}
Z(N_s=1,N_w=0,V,T)
=
\int_V d\mathbf{x}_s    
\exp 
\left[ 
    -\beta 
     \Psi_s (\mathbf{x}_{s})
\right]
\label{eq:Z1}
\end{equation}

The potential energy in the force field (FF) representation is a function only of distances between
particles and
does not depend on their absolute positions. Additionally, we consider homogeneous liquid phase.
These two facts allow us to change the coordinates of the system such that one atom of the solute
molecule is located in the origin \cite{Ben-Naim2006}. Coordinates of all solute's atoms are
written as (see Eq.~\ref{eq:bx}):
$$
\mathbf{x}_s = \{ \mathbf{r}_{s,1},\mathbf{r}_{s,2},...,\mathbf{r}_{s,n_s}\}
$$

Therefore, we may rewrite Eq.~\ref{eq:Z1}:
$$
\int_V d\mathbf{x}_s  \exp \left[ -\beta \Psi_{s} (\mathbf{x}_s)  \right] = 
\int_V d\mathbf{r}^{\prime}_{s,1} d\mathbf{r}^{\prime}_{s,2} ... d\mathbf{r}^{\prime}_{s,n_s}
\exp \left[ -\beta \Psi_{s} (0,\mathbf{r}^{\prime}_{s,2},...,\mathbf{r}^{\prime}_{s,n_s})  \right] 
= 
$$

$$
=
V \int_V d\mathbf{r}^{\prime}_{s,2} ... d\mathbf{r}^{\prime}_{s,n_s} 
\exp \left[ -\beta \Psi_{s} (0,\mathbf{r}^{\prime}_{s,2},...,\mathbf{r}^{\prime}_{s,n_s})  \right]
= 
$$

$$
=
V \int_V d\mathbf{r}_{s,2} ... d\mathbf{r}_{s,n_s} 
\exp \left[ -\beta \Psi_{s} (0,\mathbf{r}_{s,2},...,\mathbf{r}_{s,n_s})  \right]
=
$$

where we, firstly, integrated out the position of the first atom of the solute molecule which
released the volume, and, secondly, we dropped the $\prime$ marks for simplicity.

To proceed we introduce the following approximation. The bonded potentials in FF representation
do not allow atoms belonging to the same molecule to move far from each other. Therefore, the
limits of integration for the rest of solute's atoms with very high accuracy can be reduced to a
small volume around the first atom (we denote it as $V_{s}$). Please note that, since we consider a
single molecule here these do not affect the combinatorial prefactor of the canonical partition
function. Therefore we write:
$$
\int_V d\mathbf{x}_s  \exp \left[ -\beta \Psi_{s} (\mathbf{x}_s)  \right]
=
V \int_{V_{s}} ... \int_{V_{s}} d\mathbf{r}_{s,2} ... d\mathbf{r}_{s,n_s} 
\exp \left[ -\beta \Psi_{s} (0,\mathbf{r}_{s,2},...,\mathbf{r}_{s,n_s})  \right]
$$

For simplicity we will use the following
notations: $\mathbf{x}^{*}_{s} = \{ \mathbf{r}_{s,2},...,\mathbf{r}_{s,n_s} \}$, and
correspondingly, $d\mathbf{x}^{*}_{s} = d\mathbf{r}_{s,2} ...
d\mathbf{r}_{s,n_s}$. With these notations we have:
\begin{equation}
\int_V d\mathbf{x}_s  \exp \left[ -\beta \Psi_{s} (\mathbf{x}_s)  \right]
=
V \int_{V_{s}^{n_s-1}} d\mathbf{x}^{*}_s 
\exp \left[ -\beta \Psi_{s} (0,\mathbf{x}^*_{s})  \right]
= 
V 
\cdot 
q_s(T)
\label{eq:Z1_2}
\end{equation}
where we introduced new function $q_s (T)$, which is in some sense corresponds to the internal
partition function of a solute molecule in FF representation:
\begin{equation}
q_s (T) = \int_{V_s^{n_s-1}} d\mathbf{x}^*_s
  \exp \left[ -\beta \Psi_{s} (0,\mathbf{x}^*_s)  \right]
\end{equation}
However, one should note that in this definition all conformations of the molecule are taken into
account in contrast to the usual $q(T)$ definition based on vibrational, rotational, electronic,
etc. partition functions, which are defined for a single molecular conformation
\cite{McQuarrie2000}.

With the help of Eq.~\ref{eq:Z1_2} we may write the configuration integral (Eq.~\ref{eq:Z}) for
non-interacting solute molecules as:
\begin{equation}
Z_{id}(N_s,N_w=0,V,T) 
=
\int_V d\mathbf{x}_s^{N_s}    
\exp 
\left[ 
    -\beta 
     \sum_{i=1}^{N_s} \Psi (\mathbf{x}_{s,i})
\right]
=
\left(
\int_V d\mathbf{x}_s    
\exp 
\left[ 
    -\beta 
     \Psi (\mathbf{x}_{s})
\right]
\right)^{N_s}
=
\left(
V \cdot q_s(T)
\right)^{N_s}
\label{eq:Zid}
\end{equation}

And the corresponding canonical partition function (see Eq.~\ref{eq:Q}) is:
\begin{equation}
Q_{id}(N_s,N_w=0,V,T) =
 \prod_{\alpha=1}^{n_s} \frac{\Lambda_{s,\alpha}^{-3N_s}}{N_s!}  
 \cdot  
 V^{N_s} \cdot q_s^{N_s}(T)
\label{eq:Qid}
\end{equation}

The chemical potential for the ideal gas case at given $V$ is written as:
\begin{equation*}
\mu_{id} 
=
-k_BT \ln
\frac
{
Q_{id}(N_s,N_w=0,V,T)
}
{
Q_{id}(N_s-1,N_w=0,V,T)
}
=
-k_BT \ln
\frac
{
 \prod_{\alpha=1}^{n_s} \frac{\Lambda_{s,\alpha}^{-3N_s}}{N_s!}  
 \cdot  
 V^{N_s} \cdot q_s^{N_s}(T)
}
{
 \prod_{\alpha=1}^{n_s} \frac{\Lambda_{s,\alpha}^{-3(N_s-1)}}{(N_s-1)!}  
 \cdot  
 V^{(N_s-1)} \cdot q_s^{(N_s-1)}(T)
}
=
\end{equation*}
\begin{equation}
=
-k_BT \ln
\left[
\frac
{
 V  q_s(T)
}
{
 N_s^{n_s}
}
\cdot
\prod_{\alpha=1}^{n_s} \Lambda_{s,\alpha}^{-3}
\right]
\label{eq:mu_id}
\end{equation}

\subsection{Solvation free energy and excess chemical potential}


The solvation free energy (SFE) can be defined as a reversible work required to switch on the
interactions between a solute molecule and the rest \cite{Ben-Naim2006}. In NPT ensemble this
can be written as:

\begin{equation}
 \Delta G_{solv} 
 = 
 - k_BT \ln 
 \frac{\Delta(N_s,N_w,P,T,\lambda=1)}
      {\Delta(N_s,N_w,P,T,\lambda=0)}
 \label{eq:dG_solv_1}
 \end{equation}
where the $\lambda$ in the brackets indicate that the partition functions are written with the
$\lambda$-parameterized Hamiltonian (Eq.~\ref{eq:Hl}). 

In the next transformation of Eq.~\ref{eq:dG_solv_1} the $N_s^{n_s}$ factor in the
nominator appears because of the dissemination
process \cite{Ben-Naim2006}. When we write the parameterized Hamiltonian we
scale the solute-solvent interactions only for one solute molecule, which makes this solute
molecule distinguishable from the rest. This switch from assimilated and disseminated solute
molecule changes the combinatorial prefactor of the canonical partition function (Eq.~\ref{eq:Q}).
Inside a single molecule we consider atoms being physically different, however atoms of the
same type from identical molecules are physically identical. Therefore the
$N_s^{n_s}$ factor appears:

\begin{equation}
 \Delta G_{solv} 
 = 
- k_BT \ln 
 \frac{N_s^{n_s} \cdot \Delta(N_s,N_w,P,T) }
      {\int_0^\infty d\left( \frac{V}{V^\prime} \right) e^{-\beta PV} 
Q(N_s=1,N_w=0,V,T) Q(N_s-1,N_w,V,T) }
\label{eq:dGsolv}
\end{equation}
where $\Delta(N_s,N_w,P,T)$ is the partition function with non-parameterized Hamiltonian and the
canonical partition function in the denominator factorizes into canonical partition function for
the single solute molecule and the system with $N_s-1$ solvent molecules.

The denominator in Eq.~\ref{eq:dGsolv} can be further simplified: 
$$
\int_0^\infty d\left( \frac{V}{V^\prime} \right) e^{-\beta PV} Q(N_s=1,N_w=0,V,T) Q(N_s-1,N_w,V,T)
=  
$$

(please, again, note that one solute molecule is not identical to the rest and the
corresponding
combinatorial factor reduces by one:)

\begin{equation}
=
  \prod_{\alpha=1}^{n_s} \frac{\Lambda_{s,\alpha}^{-3 N_s}}{(N_s-1)!} 
  \prod_{\alpha=1}^{n_w} \frac{\Lambda_{s,\alpha}^{-3 N_w}}{N_w!} 
\cdot 
\int_0^\infty d\left( \frac{V}{V^\prime} \right)
e^{-\beta PV} Z(N_s=1,N_w=0,V,T) Z(N_s-1,N_w,V,T)
=
\label{eq:Dl0}
\end{equation}

(for a large number of molecules in the system only integration of large system volumes
contribute to the NPT partition function (Eq.~\ref{eq:Dnpt}), and therefore integration over
small volumes comparable to $V_{s}$ can be safely neglected. Thus, the
integrals over small volumes $V_s$ and, consequently, $Z(N_s=1,N_w=0,V,T)$ become independent on
total volume $V$. Therefore, using Eqs.~\ref{eq:Z1_2} we can
rewrite Eq.~\ref{eq:Dl0} as:)

\begin{equation*}
=
\left(
q_s (T)
\cdot
\prod_{\alpha=1}^{n_s} \Lambda_{s,\alpha}^{-3}
\right)
\cdot
\end{equation*}

\begin{equation}
\cdot
\left( 
  \prod_{\alpha=1}^{n_s} \frac{\Lambda_{s,\alpha}^{-3 (N_s-1)}}{(N_s-1)!}
  \prod_{\alpha=1}^{n_w} \frac{\Lambda_{s,\alpha}^{-3N_w}}{N_w!} 
  \int_0^\infty d\left( \frac{V}{V^\prime} \right) e^{-\beta PV}  
        \int_V
        d\mathbf{x}_s^{(N_s-1)} d\mathbf{x}_w^{N_w}  \cdot V \cdot Z(N_s-1,N_w,V,T)
\right)
=
\label{eq:Dl0_2}
\end{equation}

(multiplication and division of Eq.~\ref{eq:Dl0_2} by $\Delta(N_s-1,N_w,P,T)$ leads to the
following:)
$$
=
\left(
q_s (T)
\cdot
\prod_{\alpha=1}^{n_s} \Lambda_{s,\alpha}^{-3} 
\cdot
V^{*}
\right)
\cdot
\Delta(N_s-1,N_w,P,T)
$$
where $V^*$ is the average volume of the $(N_s-1,N_w,P,T)$ system. 

Finally, we write SFE (Eq.~\ref{eq:dGsolv}) as:
$$
 \Delta G_{solv} = - k_BT \ln 
\left[
 \frac
{
\Delta(N_s,N_w,P,T)
}
{
\Delta(N_s-1,N_w,P,T)
}
\cdot
\frac
{
N_s^{n_s} 
}
{
q_s (T)
\cdot
V^{*}
\cdot
\prod_{\alpha=1}^{n_s} \Lambda_{s,\alpha}^{-3} 
}
\right]
=
$$

(using Eq.~\ref{eq:mu} we get:)
$$
=
\mu
+
k_BT \ln
\frac
{
q_s (T)
\cdot
V^{*}
\cdot
\prod_{\alpha=1}^{n_s} \Lambda_{s,\alpha}^{-3} 
}
{
N_s^{n_s} 
}
=
$$

(we add $k_BT \ln \frac{V_1}{V_1}$, where $V_1$ is the mean volume of $(N_s,N_w)$ system:)
\begin{equation}
=
\mu
+
k_BT \ln 
\frac
{
q_s (T)
\cdot
V_1
\cdot
\prod_{\alpha=1}^{n_s} \Lambda_{s,\alpha}^{-3} 
}
{
N_s^{n_s} 
}
+
k_BT \ln
\frac{V^*}{V_1}
\label{eq:dGsolv_2}
\end{equation}
where the last term is the work required for one ideal gas particle to expand the volume from
$V^*$ to $V_1$. In thermodynamic limit the ratio of two volumes tends to 1 and therefore the term
vanishes. The first term is the chemical potential of the ideal gas of solute molecules
 (see Eq.~\ref{eq:mu_id}) with the average volume of $(N_s,N_w,P,T)$ system $V_1$. Therefore,
Eq.~\ref{eq:dGsolv_2}
is rewritten as:
\begin{equation}
\Delta G_{solv}
=
\mu
-
\mu^{id}
\label{eq:dGsolv_3}
\end{equation}
Eq.~\ref{eq:dGsolv_3} shows that the SFE is always equal to the excess (over ideal) chemical
potential. This also proves that the excess chemical potential is the reversible work of switching
the solute - solvent interactions, as defined in Eq.~\ref{eq:dGsolv}. Therefore, we may write:
\begin{equation}
\Delta G_{solv}
\equiv
\mu_{ex}
\label{eq:dGsolv_4}
\end{equation}

We would like to note two things here. Firstly, SFE equals to the excess chemical potential not only
at infinite dillution, as it is sometimes wrongly believed. 

Secondly, there is a wrong statement (at least for the present case of the FF models of molecules)
in the book of Ben-Naim (Ref.~\cite{Ben-Naim2006}, page 200) that "... only when $q_s$ is
unaffected by the solvation process, $\mu_{ex}$ becomes identical with the solvation Gibbs
energy ...". In our derivation of Eq.~\ref{eq:dGsolv_4} we explicitly considered the case when
internal degrees of freedom of molecules, represented by sets of interaction sites, are coupled to
other degrees of freedom.

\subsection{Kirkwood charging formula}

In order to make the forthcoming derivations simpler, from now we explicitly consider the
case of infinite diluted ssoluteion: $N_s=1$. The point here is that we will express the excess
chemical potential of solute via the
particle density distributions. Therefore, considering many solute molecules in the system
will require to introduce the solute-solute density distribution, which will unnecessarily
complicate the derivations. Note, that all the forthcoming derivation can be straightforwardly
extended to the case of multicomponent solvent (see Ref.~\cite{Sakuraba2014}).

Let us define the excess chemical potential
for the system with parameterized Hamiltonian (Eq.~\ref{eq:Hl}) at a certain $\lambda$ value. With
Eqs.~\ref{eq:dGsolv} and \ref{eq:dGsolv_4} we get:

\begin{equation*}
 \mu_{ex,\lambda} = - k_BT \ln 
 \frac{\Delta(N_s,N_w,P,T,\lambda)}
      {\Delta(N_s,N_w,P,T,\lambda=0)}
\end{equation*}

Since the denominator does not depend on $\lambda$ one can write:
\begin{equation*}
 \frac{\partial \mu_{ex,\lambda}}
      {\partial \lambda}
  = - k_BT 
  \frac
{
  \int_0^\infty d\left( \frac{V}{V^\prime} \right) e^{-\beta PV} 
  \int_V d\mathbf{x}_s d\mathbf{x}_w^{N_w} 
  \frac{
        \partial U_\lambda
       }
       {
        \partial \lambda
       }
       \exp \left[ 
                  -\beta U_\lambda (\mathbf{x}_s,\mathbf{x}_w^{N_w})
            \right]
}
{
\Delta(N_s,N_w,P,T,\lambda)
}
= 
       \left\langle 
          \frac{\partial U_\lambda}
               {\partial \lambda}  
       \right\rangle _ {\lambda} 
\end{equation*}

With the explicit form of the potential function (Eq.~\ref{eq:Ul}) we have:
\begin{equation*}
 \frac{\partial \mu_{ex,\lambda}}
      {\partial \lambda}
  = 
  \left\langle  
  \sum_{i=1}^{N_w} \frac{\partial u_{sw,\lambda}
(\mathbf{x}_s,\mathbf{x}_{w,i})}{\partial \lambda}  
\right\rangle _ {\lambda} 
=
  \left\langle  
    \int_{-\infty}^{+\infty} d\mathbf{x}_s^\prime d\mathbf{x}_{w}^\prime  
    \frac{\partial u_{sw,\lambda} (\mathbf{x}_s^\prime,\mathbf{x}_{w}^\prime)}
         {\partial \lambda}   
    \sum_{i=1}^{N_w}  \delta(\mathbf{x}_s - \mathbf{x}_s^\prime)
    \delta(\mathbf{x}_{w,i} - \mathbf{x}_w^\prime) 
  \right\rangle _ {\lambda}
\end{equation*}
where $\left\langle  \cdot  \right\rangle _ {\lambda}$ denotes ensemble average in
isothermo-isobaric condition at given $\lambda$.
We can change order of integration and take out the derivative from the ensemble average:

\begin{equation}
 \frac{\partial \mu_{ex,\lambda}}
      {\partial \lambda}
=
\int_{-\infty}^{+\infty} d\mathbf{x}_s^\prime d\mathbf{x}_{w}^\prime  \frac{\partial
u_{sw,\lambda} (\mathbf{x}_s^\prime,\mathbf{x}_{w}^\prime)}{\partial \lambda} 
\left\langle   
\sum_{i=1}^{N_w}  \delta(\mathbf{x}_s - \mathbf{x}_s^\prime) \delta(\mathbf{x}_{w,i} -
\mathbf{x}_w^\prime)  
\right\rangle _ {\lambda}
\label{eq:Kcf_0}
\end{equation}

In the right hand side there is the pair solute-solvent density distribution in
NPT-ensemble by definition (see e.g. Eq. (2.5.13) of Ref.~\cite{Hansen1991}, but mind that for
density distributions of non-identical particles the sum should include terms with $i=j$):

$$
 \frac{\partial \mu_{ex,\lambda}}
      {\partial \lambda}
=
\int d\mathbf{x}_s^\prime d\mathbf{x}_{w}^\prime  \frac{\partial
u_{sw,\lambda} (\mathbf{x}_s^\prime,\mathbf{x}_{w}^\prime)}{\partial \lambda} 
\rho_{sw,\lambda} (\mathbf{x}_s^\prime, \mathbf{x}_w^\prime)
$$

Finally, the excess chemical potential can be written as an integral over lambda:

\begin{equation}
\mu_{ex} = \int_0^1 d\lambda  \frac{\partial \mu_{ex,\lambda}}
                                   {\partial \lambda}
=
\int_0^1 d\lambda 
\int_{-\infty}^{+\infty} d\mathbf{x}_s^\prime d\mathbf{x}_{w}^\prime  \frac{\partial
u_{sw,\lambda} (\mathbf{x}_s^\prime,\mathbf{x}_{w}^\prime)}{\partial \lambda} 
\rho_{sw,\lambda} (\mathbf{x}_s^\prime, \mathbf{x}_w^\prime)
\label{eq:Kcf}
\end{equation}

Note, that Eq.~\ref{eq:Kcf} is different from Eq. (3) of Ref.~\cite{Sakuraba2014}, where the
delta function for the solute degrees of freedom is omitted (by mistake?).

\section{Energy representation (ER)}

\subsection{Basic definitions in ER}

\paragraph{Collective coordinate.}

We define a new collective coordinate which is the interaction energy between a solute molcule and a
solvent molecule: $\epsilon$. We make this coordinate $\lambda$-independent, such that this
coordinate is
calculated with the solute-solvent potential at full coupling $v_{sw}(\mathbf{x}_s,\mathbf{x}_w)$,
irrespective of the ensemble and Hamiltonian which were used to generate this configuration:
\begin{equation}
v_{sw}(\mathbf{x}_s,\mathbf{x}_w) 
\equiv 
u_{sw,\lambda=1}(\mathbf{x}_s,\mathbf{x}_w)
\label{eq:v}
\end{equation}

\paragraph{Microscopic density.}
For a single configuration of the system the microscopic density in energy representation can be
written as:
\begin{equation}
 \hat{\rho}^e_{sw} (\epsilon) = \sum_{i=1}^{N_w} \delta \left( v_{sw}(\mathbf{x}_s,\mathbf{x}_w) -
\epsilon \right)
=
\label{eq:rhoehat}
\end{equation}

\begin{equation}
= 
\int_{-\infty}^{+\infty} d\mathbf{x}^\prime_{s} d\mathbf{x}^\prime_{w} 
\delta( v_{sw}(\mathbf{x}^\prime_s,\mathbf{x}^\prime_{w}) - \epsilon ) 
\sum_{i=1}^{N_w} 
\delta( \mathbf{x}^\prime_s - \mathbf{x}_s)
\delta( \mathbf{x}^\prime_{w} - \mathbf{x}_{w,i})
=
\label{eq:rhoehat_2}
\end{equation}

\begin{equation}
= 
\int_{-\infty}^{+\infty} d\mathbf{x}_{s} d\mathbf{x}_{w} 
\delta( v_{sw}(\mathbf{x}_s,\mathbf{x}_{w}) - \epsilon ) 
\hat{\rho}_{sw} (\mathbf{x}_s,\mathbf{x}_{w})
\label{eq:rhoehat_3}
\end{equation}

\paragraph{Potential in ER.}
The potential in energy representation can be written as:
\begin{equation}
 u^e_{sw,\lambda} (\epsilon) 
 = 
 \int_{-\infty}^{+\infty} d\mathbf{x}_s d\mathbf{x}_w   
 \delta \left(
          v_{sw}(\mathbf{x}_s,\mathbf{x}_w) - \epsilon 
       \right) 
 u_{sw,\lambda}(\mathbf{x}_s,\mathbf{x}_w)
\label{eq:ue}
\end{equation}

It is important for the following derivation that we choose the lambda path in such a way that
$u_{sw,\lambda} (\mathbf{x}_s,\mathbf{x}_w)$ is constant on each equi-energy surface of $v_{sw}
(\mathbf{x}_s,\mathbf{x}_w)$. This can be achieved, for instance, when $u_{sw,\lambda}
(\mathbf{x}_s,\mathbf{x}_w) = \lambda v_{sw} (\mathbf{x}_s,\mathbf{x}_w)$. With this restriction of
the $u_{sw,\lambda}$ potential we can write the following identity:

\begin{equation}
 u_{sw,\lambda}(\mathbf{x}_s,\mathbf{x}_w) = \int_{-\infty}^{+\infty} d\epsilon 
 \cdot
 \delta \left(
v_{sw}(\mathbf{x}_s,\mathbf{x}_w) - \epsilon \right) u^e_{sw,\lambda} (\epsilon)
\label{eq:ux}
\end{equation}

Taking the partial derivative of the both sides of equation we obtain the formula, which will be
used later on:
\begin{equation}
 \frac{\partial u_{sw,\lambda}(\mathbf{x}_s,\mathbf{x}_w)}{\partial \lambda} 
 = 
 \int_{-\infty}^{+\infty} d\epsilon \cdot \delta \left(
v_{sw}(\mathbf{x}_s,\mathbf{x}_w) - \epsilon \right) 
\frac{\partial u^e_{sw,\lambda} (\epsilon)}
     {\partial \lambda}
\label{eq:dlux}
\end{equation}

\paragraph{Solute-solvent density distribution in ER.}
Solute-solvent density distribution in NPT ensemble is written as:
$$
\rho^e_{sw,\lambda} (\epsilon) 
= 
\left\langle   
  \hat{\rho}(\epsilon)  
\right\rangle_{\lambda}
$$

Using the definition of microscopic density in ER (Eq.~\ref{eq:rhoehat_3}) and writing explicitly
its ensemble average we get: 

\begin{equation}
\rho^e_{sw,\lambda} (\epsilon)
=
\frac{
        \int_0^\infty d\left( \frac{V}{V^\prime} \right) e^{-\beta PV} 
        \int_V d\mathbf{x}_s d\mathbf{x}_w^{N_w}  
         \left[   
                \int_{-\infty}^{+\infty}  d\mathbf{x}^\prime_{s} d\mathbf{x}^\prime_{w} 
\delta( v_{sw}(\mathbf{x}^\prime_s,\mathbf{x}^\prime_{w}) - \epsilon ) 
\hat{\rho}_{sw} (\mathbf{x}_s^\prime,\mathbf{x}_{w}^\prime)
         \right]
        \exp \left[ -\beta U_\lambda (\mathbf{x}_s,\mathbf{x}_w^{N_w}) \right]
     }
     {
        \int_0^\infty d\left( \frac{V}{V^\prime} \right) e^{-\beta PV} 
        \int_V d\mathbf{x}_s d\mathbf{x}_w^{N_w}  
        \exp \left[ -\beta U_\lambda (\mathbf{x}_s,\mathbf{x}_w^{N_w}) \right]
     }
\label{eq:rhoe2}
\end{equation}

Change of the integration order:

$$
\rho^e_{sw,\lambda} (\epsilon)
=
        \int_{-\infty}^{+\infty} d\mathbf{x}^\prime_{s} d\mathbf{x}^\prime_{w} 
\delta( v_{sw}(\mathbf{x}^\prime_s,\mathbf{x}^\prime_{w}) - \epsilon )
\frac{
        \int_0^\infty d\left( \frac{V}{V^\prime} \right) e^{-\beta PV} 
        \int_V d\mathbf{x}_s d\mathbf{x}_w^{N_w}  
         \left[  
\hat{\rho}_{sw} (\mathbf{x}_s^\prime,\mathbf{x}_{w}^\prime)
         \right]
        \exp \left[ -\beta U_\lambda (\mathbf{x}_s,\mathbf{x}_w^{N_w}) \right]
     }
     {
        \int_0^\infty d\left( \frac{V}{V^\prime} \right) e^{-\beta PV} 
        \int_V d\mathbf{x}_s d\mathbf{x}_w^{N_w}  
        \exp \left[ -\beta U_\lambda (\mathbf{x}_s,\mathbf{x}_w^{N_w}) \right]
     }
$$

The ratio gives us the definition of the solute-solvent density distribution (see Comment after
Eq.~\ref{eq:Kcf_0}):
\begin{equation}
\rho^e_{sw,\lambda} (\epsilon)
=
        \int_{-\infty}^{+\infty} d\mathbf{x}^\prime_{s} d\mathbf{x}^\prime_{w} 
\delta( v_{sw}(\mathbf{x}^\prime_s,\mathbf{x}^\prime_{w}) - \epsilon )
\rho_{sw,\lambda}(\mathbf{x}^\prime_s,\mathbf{x}^\prime_{w})
\label{eq:rhoe4}
\end{equation}

\subsection{Kirkwood charging formula in energy representation}\label{sec:KCF_ER}

\subsubsection{Kirkwood charging formula via density distribution}

Let us obtain the charging formula in energy representation. We start from coordinate
representation (Eq.~\ref{eq:Kcf}):

\begin{equation*}
\mu_{ex} = \int_0^1 d\lambda  
\int_{-\infty}^{+\infty} d\mathbf{x}_s^\prime d\mathbf{x}_{w}^\prime  \frac{\partial
u_{sw,\lambda} (\mathbf{x}_s^\prime,\mathbf{x}_{w}^\prime)}{\partial \lambda} 
\rho_{sw,\lambda} (\mathbf{x}_s^\prime, \mathbf{x}_w^\prime)
=
\end{equation*}

Using Eq.~\ref{eq:dlux} we obtain:

$$
\mu_{ex}
=
\int_0^1 d\lambda  
\int_{-\infty}^{+\infty} d\mathbf{x}_s^\prime d\mathbf{x}_{w}^\prime 
\left[
\int_{-\infty}^{+\infty} d\epsilon \delta \left(
v_{sw}(\mathbf{x}_s^\prime, \mathbf{x}_w^\prime) - \epsilon \right) 
\frac{\partial u^e_{sw,\lambda} (\epsilon)}
     {\partial \lambda}
\right]
     \rho_{sw,\lambda} (\mathbf{x}_s^\prime, \mathbf{x}_w^\prime)
$$

We change the integration order:
$$
\mu_{ex}
=
\int_0^1 d\lambda  
\int_{-\infty}^{+\infty} d\epsilon
\frac{\partial u^e_{sw,\lambda} (\epsilon)}
     {\partial \lambda}
\left[
 \int_{-\infty}^{+\infty} d\mathbf{x}_s^\prime d\mathbf{x}_{w}^\prime
 \delta \left(
v_{sw}(\mathbf{x}_s^\prime,\mathbf{x}_w^\prime) - \epsilon \right) 
     \rho_{sw,\lambda} (\mathbf{x}_s^\prime, \mathbf{x}_w^\prime)
\right]
$$

We use the relation Eq.~\ref{eq:rhoe4} to obtain:
\begin{equation}
\mu_{ex}
=
\int_0^1 d\lambda  
\int_{-\infty}^{+\infty} d\epsilon
\frac{\partial u^e_{sw,\lambda} (\epsilon)}
     {\partial \lambda}
\rho^e_{sw,\lambda} (\epsilon)
\label{eq:Kcf_e}
\end{equation}

Eq.~\ref{eq:Kcf_e} is the Kirkwood's charging formula in energy representation.

\subsubsection{Indirect part of potential of mean force (IPMF)}

We can introduce an auxiliary function $w^e_{sw,\lambda}(\epsilon)$, which is an analogue of the
indirect part
of potential of mean force in coordinate representation:
\begin{equation}
\rho^e_{sw,\lambda}(\epsilon) 
=
\rho^e_{sw,\lambda=0}(\epsilon) 
\cdot
\exp 
\left[ 
  -\beta \left( 
            u^e_{sw,\lambda} (\epsilon) + w^e_{sw,\lambda} (\epsilon)
         \right)
\right]
\label{eq:we}
\end{equation}

The potential then can be rewritten as:
\begin{equation}
u^e_{sw,\lambda} (\epsilon) 
=
- k_BT \ln 
\frac
{
\rho^e_{sw,\lambda}(\epsilon) 
}
{
\rho^e_{sw,\lambda=0}(\epsilon)
}
- w^e_{sw,\lambda} (\epsilon)
\label{eq:uwe}
\end{equation}

\subsubsection{Kirkwood charging formula via IPMF}

Let us rewrite the Kirkwood's charging formula (Eq.~\ref{eq:Kcf_e}) via $w^e_{sw,\lambda}$:
$$
\mu_{ex}
= 
\int_0^1 d\lambda  
\int_{-\infty}^{+\infty} d\epsilon
\frac{\partial u^e_{sw,\lambda} (\epsilon)}
     {\partial \lambda}
\rho^e_{sw,\lambda} (\epsilon)
$$

Change of the integration order:
$$
\mu_{ex}
= 
\int_{-\infty}^{+\infty} d\epsilon
\int_0^1 d\lambda  
\frac{\partial u^e_{sw,\lambda} (\epsilon)}
     {\partial \lambda}
\rho^e_{sw,\lambda} (\epsilon)
$$

Integration by parts for the inner integral:
$$
\mu_{ex}
= 
\int_{-\infty}^{+\infty} d\epsilon
\left[
 \rho^e_{sw,\lambda=1} (\epsilon) u^e_{sw,\lambda=1} (\epsilon)
- \int_0^1 d\lambda  
\frac{\partial \rho^e_{sw,\lambda} (\epsilon) }
     {\partial \lambda}
u^e_{sw,\lambda} (\epsilon)
\right]
$$

Change of the integration order back. Mind that $u^e_{sw,\lambda=1} (\epsilon) = v^e_{sw}
(\epsilon) = \epsilon$ according to the definition (Eq.~\ref{eq:v} and Eq.~\ref{eq:ue}):
\begin{equation}
\mu_{ex}
= 
\int_{-\infty}^{+\infty} d\epsilon
\rho^e_{sw,\lambda=1} (\epsilon) \epsilon
- 
\int_0^1 d\lambda  
\int_{-\infty}^{+\infty} d\epsilon
\frac{\partial \rho^e_{sw,\lambda} (\epsilon) }
     {\partial \lambda}
u^e_{sw,\lambda} (\epsilon)
\label{eq:Kcf_e_2}
\end{equation}

Let us denote the last term as a functional of the potential and the solute-solvent density
distribution:
\begin{equation}
\mathcal{F} [ \rho^e_{sw,\lambda}(\epsilon), u^e_{sw,\lambda} (\epsilon)]
=
\int_0^1 d\lambda  
\int_{-\infty}^{+\infty} d\epsilon
\frac{\partial \rho^e_{sw,\lambda} (\epsilon) }
     {\partial \lambda}
u^e_{sw,\lambda} (\epsilon)
\label{eq:F}
\end{equation}


The functional can be written via IPMF. 
Here and after, we use the following simplified notations: 
$$
\rho^e_{sw,0} \equiv \rho^e_{sw,\lambda=0}
$$
$$
\rho^e_{sw} \equiv \rho^e_{sw,\lambda=1}
$$
Similar notations are adopted for other functions. 

Using Eq.~\ref{eq:uwe} and changing the integration order we obtain from Eq.~\ref{eq:F}:
\begin{equation}
\mathcal{F} [ \rho^e_{sw,\lambda}(\epsilon), u^e_{sw,\lambda} (\epsilon)]
= 
\int_{-\infty}^{+\infty} d\epsilon
\int_0^1 d\lambda 
\frac{\partial \rho^e_{sw,\lambda} (\epsilon) }
     {\partial \lambda}
\left(
    - k_BT \ln \frac{\rho^e_{sw,\lambda}(\epsilon)}{\rho^e_{sw,0}(\epsilon)}
    - w^e_{sw,\lambda} (\epsilon)
\right)
\label{eq:F1}
\end{equation}

The first integral in Eq.~\ref{eq:F1} can be taken analytically by parts:
$$
\int_0^1 d\lambda 
\frac{\partial \rho^e_{sw,\lambda}(\epsilon) }
     {\partial \lambda}
    \ln \frac{\rho^e_{sw,\lambda}(\epsilon)}{\rho^e_{sw,0}(\epsilon)}
=
  \left.
       \rho^e_{sw,\lambda}(\epsilon)
       \ln \frac{\rho^e_{sw,\lambda}(\epsilon) }{\rho^e_{sw,0}(\epsilon) }
  \right\vert_0^1
  - \int_0^1 d\lambda \frac{\rho^e_{sw,\lambda}(\epsilon)}{\rho^e_{sw,\lambda}(\epsilon)} 
  \frac{\partial \rho^e_{sw,\lambda}(\epsilon) }
       {\partial \lambda}
=
$$

\begin{equation}
=
   \rho^e_{sw}(\epsilon) \ln \frac{\rho^e_{sw}(\epsilon)}{\rho^e_{sw,0}(\epsilon)}
  - \left(
	\rho^e_{sw}(\epsilon) - \rho^e_{sw,0}(\epsilon) 
    \right) 
\label{eq:I1}
\end{equation}

Therefore, we rewrite Eq.~\ref{eq:F1} using Eq.~\ref{eq:I1} as:
$$
\mathcal{F} [ \rho^e_{sw,\lambda}(\epsilon), u^e_{sw,\lambda} (\epsilon)] =
$$
\begin{equation*}
= 
\int_{-\infty}^{+\infty} d\epsilon
\left[
- k_BT 
\left(
   \rho^e_{sw}(\epsilon) \ln \frac{\rho^e_{sw}(\epsilon)}{\rho^e_{sw,0}(\epsilon)}
  - \left(
	\rho^e_{sw}(\epsilon) - \rho^e_{sw,0}(\epsilon) 
    \right) 
\right)
+
\int_0^1 d\lambda
\frac{\partial \rho^e_{sw,\lambda}(\epsilon) }
     {\partial \lambda}
\left(
    - w^e_{sw,\lambda}(\epsilon)
\right)
\right]
\end{equation*}

Regrouping the terms we get:

\begin{equation}
\mathcal{F} [ \rho^e_{sw,\lambda}(\epsilon), u^e_{sw,\lambda} (\epsilon)]
= 
k_BT 
\int_{-\infty}^{+\infty} d\epsilon
\left[
   \left(
	\rho^e_{sw} (\epsilon)- \rho^e_{sw,0}(\epsilon) 
    \right)
  - \rho^e_{sw}(\epsilon) \ln \frac{\rho^e_{sw}(\epsilon)}{\rho^e_{sw,0}(\epsilon)}
-
\beta
\int_0^1 d\lambda
\frac{\partial \rho^e_{sw,\lambda} (\epsilon) }
     {\partial \lambda}
    w^e_{sw,\lambda} (\epsilon)
\right]
\label{eq:F2}
\end{equation}

This expression can be further simplified if we choose the $\lambda$-dependence of the potential
such that the density distribution is a linear function of $\lambda$:
\begin{equation}
\rho^e_{sw,\lambda}(\epsilon) = \lambda \rho^e_{sw}(\epsilon) + (1-\lambda) \rho^e_{sw,0}(\epsilon)
\label{eq:rhoel}
\end{equation}

With this restriction (Eq.~\ref{eq:rhoel}) the $\lambda$-derivative is:
$$
\frac{\partial \rho^e_{sw,\lambda} (\epsilon) }
     {\partial \lambda}
=
(\rho^e_{sw}(\epsilon) - \rho^e_{sw,0}(\epsilon))
$$

and the functional (Eq.~\ref{eq:F2}) becomes:
$$
\mathcal{F} [ \rho^e_{sw,\lambda}(\epsilon), u^e_{sw,\lambda} (\epsilon)] =
$$
\begin{equation}
= 
k_BT 
\int_{-\infty}^{+\infty} d\epsilon
\left[
   \left(
	\rho^e_{sw} (\epsilon)- \rho^e_{sw,0}(\epsilon) 
    \right)
  - \rho^e_{sw}(\epsilon) \ln \frac{\rho^e_{sw}(\epsilon)}{\rho^e_{sw,0}(\epsilon)}
-
\beta
(\rho^e_{sw}(\epsilon) - \rho^e_{sw,0}(\epsilon))
\int_0^1 d\lambda
    w^e_{sw,\lambda} (\epsilon)
\right]
\label{eq:F3}
\end{equation}

Finally, the excess chemical potential (Eq.~\ref{eq:Kcf_e}) reads:
\begin{equation}
\mu_{ex} [ \rho^e_{sw,\lambda}(\epsilon), u^e_{sw,\lambda} (\epsilon)]
= 
\int_{-\infty}^{+\infty} d\epsilon
\rho^e_{sw} (\epsilon) \epsilon
- 
\mathcal{F} [\rho^e_{sw,\lambda}(\epsilon), u^e_{sw,\lambda} (\epsilon)]
\label{eq:Kcf_e_2}
\end{equation}

\subsubsection{Density functional}

For further derivations we would like to consider the functional $\mathcal{F}$ as a unique
functional of $\rho^e_{sw,\lambda}(\epsilon)$. This can be the case if the solute-solvent
interaction potential is a unique functional of $\rho^e_{sw,\lambda}(\epsilon)$. This implies
that there should be only one $u^e_{sw,\lambda}(\epsilon)$ to which a given
$\rho^e_{sw,\lambda}(\epsilon)$ corresponds. Both in coordinate and energy representation it is not
the case if we consider 
ensembles where the number of particles is fixed
\cite{Matubayasi2003,Sakuraba2014,White2001,Hernando2002}. This can be easily seen from the
definition of $\rho^e_{sw,\lambda}(\epsilon)$ (Eq.~\ref{eq:rhoe2} and Eq.~\ref{eq:Ul}): if one adds
a constant to the solute-solvent interaction potential $u_{sw,\lambda}$ the resulting
$\rho^e_{sw,\lambda}$ function does not change (mind, that there is a one-to-one correspondence
between the
potential in energy and coordinate representations (Eqs.~\ref{eq:ux} and \ref{eq:ue})). The lack of
the one-to-one correspondence
between $\rho$ and $u$ results to the fact that the density-density correlation matrix is not
invertible and has a singular eigenvalue \cite{Matubayasi2003,Sakuraba2014,White2001}. 

Matubayasi proposed a way how to retain the one-to-one $\rho$ - $u$ correspondence by introducing
additional condition based on the physical sense. Firstly, he showed that the potentials
giving
different density profiles can differ from each other only by an additive constant
(see Appendix of Ref.~\cite{Sakuraba2014} and Ref.~\cite{Matubayasi2003}). Secondly, he set the
additive constant to ensure that
the chemical potential is an intensive property of the system. This can be achieved by ensuring
that the solute-solvent pair potential reaches zero when particle separation tends to infinity.

With this approach a one-to-one correspondence between $u_{sw,\lambda}(\epsilon)$ and
$\rho_{sw,\lambda}(\epsilon)$ achieved both in coordinate and energy representation.
This allows us to consider the potential $u^e_{sw,\lambda}(\epsilon)$ as a functional of
$\rho^e_{sw,\lambda}(\epsilon)$ in a fixed-N ensemble and use the functional calculus to obtain
approximate free energy functionals.

Therefore, the excess chemical potential (Eq.~\ref{eq:Kcf_e_2}) can be written as a density
functional of the solute-solvent density distribution:
\begin{equation}
\mu_{ex} [ \rho^e_{sw,\lambda}(\epsilon) ]
= 
\int_{-\infty}^{+\infty} d\epsilon
\rho^e_{sw} (\epsilon) \epsilon
- 
\mathcal{F} [ \rho^e_{sw,\lambda}(\epsilon) ]
\label{eq:Kcf_e_3}
\end{equation}

\subsection{Approximate free energy functional.}

The exact free energy functional (Eq.~\ref{eq:F3}) contains the term which depends on $\lambda$. To
eliminate the $\lambda$-dependence we apply the Percus's method of functional expansion to obtain
approximate functionals.

\subsubsection{Hypernetted chain (HNC) - like approximation.}

Following Percus \cite{Percus1962} we obtain the HNC-like approximation by expanding the following
functional in powers of density fluctuations $\rho^e_{sw}(\epsilon^\prime) -
\rho^e_{sw,0}(\epsilon^\prime)$:
\begin{equation}
 \ln \rho^e_{sw}(\epsilon) + \beta u^e_{sw}(\epsilon) 
 \approx
 \ln \rho^e_{sw,0}(\epsilon) 
 + 
 \int_{-\infty}^{+\infty} d\epsilon^\prime \cdot 
 \left( 
  \rho^e_{sw}(\epsilon^\prime) - \rho^e_{sw,0}(\epsilon^\prime)
 \right)
 \cdot
 \left.
 \frac
      {\delta \left[ \ln \rho^e_{sw}(\epsilon) 
                     + \beta u^e_{sw}(\epsilon)
              \right] 
      }
      {\delta \rho^e_{sw}(\epsilon^\prime) }
  \right|_{\rho^e_{sw}(\epsilon^\prime) = \rho^e_{sw,0}(\epsilon^\prime)}      
 \label{eq:HNC}
\end{equation}

With the help of Eq.~\ref{eq:uwe} we rewrite the left hand side of Eq.~\ref{eq:HNC} via IPMF.
Therefore, IPMF in HNC-like approximation can be
written as:
\begin{equation}
 w^{e,HNC}_{sw}(\epsilon)
 =
 -k_BT
 \int_{-\infty}^{+\infty} d\epsilon^\prime \cdot 
 \left( 
  \rho^e_{sw}(\epsilon^\prime) - \rho^e_{sw,0}(\epsilon^\prime)
 \right)
 \cdot
 \left[
 \frac
      {\delta (\epsilon - \epsilon^\prime) }
      {\rho^e_{sw,0}(\epsilon) }
    +
    \beta
    \left.
     \frac
      {\delta u^e_{sw}(\epsilon)  }
      {\delta \rho^e_{sw}(\epsilon^\prime) }
    \right|_{\rho^e_{sw}(\epsilon^\prime) = \rho^e_{sw,0}(\epsilon^\prime)}    
  \right]
 \label{eq:wHNC}
\end{equation}

Let us show that the functional derivative in Eq.~\ref{eq:wHNC} is a functional inverse of the
density-density correlation function. For that we start from the definition of solute-solvent
distribution function at full solute coupling:
$$
\rho^e_{sw} (\epsilon) = 
\left\langle   \hat{\rho}(\epsilon)  \right\rangle_{\lambda=1}
=
$$


\begin{equation}
=
\frac{
        \int_0^\infty d\left( \frac{V}{V^\prime} \right) e^{-\beta PV} 
        \int_V d\mathbf{x}_s d\mathbf{x}_w^{N_w}  
	\hat{\rho}^e (\epsilon)
        e^ { 
             -\beta  U (\mathbf{x}_s,\mathbf{x}_w^{N_w} ) 
           }
     }
     {
        \int_0^\infty d\left( \frac{V}{V^\prime} \right) e^{-\beta PV} 
        \int_V d\mathbf{x}_s d\mathbf{x}_w^{N_w}  
        e^ { 
             -\beta  U (\mathbf{x}_s,\mathbf{x}_w^{N_w} ) 
           }
     }
\label{eq:rhoe3}
\end{equation}

Let us denote nominator of Eq.~\ref{eq:rhoe3} as $f$ and denominator as $g$. Then, find the
functional derivative of distribution function with respect to solute-solvent potential:
\begin{equation}
 \frac{\delta \rho^e_{sw}(\epsilon)}
      {\delta u^e_{sw}(\epsilon^{\prime\prime})}
 = 
 \frac
 {
 \frac{\delta f }
      {\delta u^e_{sw} }
 }
 {
  g
 }
 -
 \frac
 {
 \frac{\delta g}
      {\delta u^e_{sw}}
 }
 {
  g
 }
 \cdot
 \frac
 {
  f
 }
 {
  g
 }
 \label{eq:fg}
\end{equation}

Both in $f$ and $g$ only potential energy $U$ depends on
$u^e_{sw}$. To write its explicit dependence on $u^e_{sw}$ we use the relation between the
solute-solvent interaction potentials in coordinate and energy representations (Eq.~\ref{eq:ux}):
\begin{equation*}
 U (\mathbf{x}_s,\mathbf{x}_w^{N_w} ) 
 = 
 \Psi (\mathbf{x}_s) + \sum_{i=1}^{N_w} u_{sw}
(\mathbf{x}_s,\mathbf{x}_{w,i}) + U_{ww} (\mathbf{x}_{w}^{N_w})
=
\end{equation*}

\begin{equation}
 = 
 \Psi (\mathbf{x}_s) 
 + 
 \sum_{i=1}^{N_w} 
 \int_{-\infty}^{+\infty} d\epsilon^{\prime\prime} 
 \cdot \delta ( v_{sw}(\mathbf{x}_s,\mathbf{x}_{w,i} ) - \epsilon^{\prime\prime} ) u^e_{sw}
 (\epsilon^{\prime\prime})
 + 
 U_{ww} (\mathbf{x}_{w}^{N_w})
\end{equation}

Therefore, we find the following derivative which will be used in later derivations:
\begin{equation*}
 \frac{\delta \left[ e^{-\beta U } \right] }
      {\delta u^e_{sw} (\epsilon^{\prime}) }
 =
 -\beta e^{-\beta U } 
 \sum_{i=1}^{N_w}
 \int_{-\infty}^{+\infty} d\epsilon^{\prime\prime} 
 \cdot \delta ( v_{sw}(\mathbf{x}_s,\mathbf{x}_{w,i} ) - \epsilon^{\prime\prime} ) 
       \delta ( \epsilon^{\prime\prime} - \epsilon^{\prime} )
 =
\end{equation*}

\begin{equation}
 =
 -\beta e^{-\beta U } 
 \sum_{i=1}^{N_w}
 \delta ( v_{sw}(\mathbf{x}_s,\mathbf{x}_{w,i} ) - \epsilon^{\prime} )
 =
  -\beta e^{-\beta U } 
 \hat{\rho}^e_{sw} (\epsilon^{\prime})
\label{eq:deUdu}
\end{equation}

where we used Eq.~\ref{eq:ue}. 

With this relation (Eq.~\ref{eq:deUdu}) the first term in
Eq.~\ref{eq:fg} then can be
written as:
\begin{equation}
 \frac
 {
 \frac{\delta f }
      {\delta u^e_{sw} }
 }
 {
  g
 }
 =
 -\beta \left\langle
          \hat{\rho}^e_{sw} (\epsilon)
          \hat{\rho}^e_{sw} (\epsilon^{\prime})
        \right\rangle_{u_{sw}}
\end{equation}
where  $\left\langle \cdot \right\rangle_{u_{sw}}$ denotes the ensemble average with the
Hamiltonian where the solute-solvent interaction potential is $u_{sw}$.

Also, with relation Eq.~\ref{eq:deUdu} we see that $g^\prime = -\beta f$. 
With this Eq.~\ref{eq:deUdu} is written as:
\begin{equation}
 \frac{\delta \rho^e_{sw}(\epsilon)}
      {\delta u^e_{sw}(\epsilon^{\prime})}
 =  
 -\beta 
     \left[
        \left\langle
          \hat{\rho}^e_{sw} (\epsilon)
          \hat{\rho}^e_{sw} (\epsilon^{\prime})
        \right\rangle_{u_{sw}}
        -
        \left\langle
          \hat{\rho}^e_{sw} (\epsilon)
        \right\rangle_{u_{sw}}
        \left\langle
          \hat{\rho}^e_{sw} (\epsilon^{\prime})
        \right\rangle_{u_{sw}}
     \right]
 \label{eq:chi}
\end{equation}

Which equivalently can be written as:
\begin{equation}
 \frac{\delta \rho^e_{sw}(\epsilon)}
      {\delta u^e_{sw}(\epsilon^{\prime})}
 =  
 -\beta 
     \left[
        \rho^{e}_{sww} (\epsilon,\epsilon^\prime)
        +
        \rho^e_{sw} (\epsilon) \delta ( \epsilon - \epsilon^\prime )
        -
        \rho^e_{sw} (\epsilon) 
        \rho^e_{sw} (\epsilon^\prime)
     \right]
 =
 -\beta
 \chi^e_{sww} (\epsilon,\epsilon^\prime)
 \label{eq:chi2}
\end{equation}
where $\chi^e_{sww} (\epsilon,\epsilon^\prime)$ is the density-density correlation function, and
$\rho^{e}_{sww} (\epsilon,\epsilon^\prime)$ is the three molecule distribution density distribution
defined by analogy to the two molecule density distribution in coordinate representation (see
Eq. (2.5.13) of Ref.~\cite{Hansen1991}) as:
\begin{equation}
 \rho^{e}_{sww} (\epsilon,\epsilon^\prime)
 =
 \left\langle
    \sum_{i=1}^{N_w} \sum_{j \neq i} 
	\delta ( v (\mathbf{x}_s,\mathbf{x}_{w,i}) - \epsilon )
	\delta ( v (\mathbf{x}_s,\mathbf{x}_{w,j}) - \epsilon^\prime )
  \right\rangle_{u_{sw}}
 \label{eq:chi3}
\end{equation}

From Eqs.~\ref{eq:chi} and \ref{eq:chi2} we obtain:
\begin{equation}
 \frac{\delta u^e_{sw}(\epsilon) }
      {\delta \rho^e_{sw}(\epsilon^{\prime}) }
 =
 \left(
 \frac{\delta \rho^e_{sw}(\epsilon) }
      {\delta u^e_{sw}(\epsilon^{\prime}) } 
 \right)^{-1}
 =
 -k_BT 
     \left( \chi^e_{sww} \right)^{-1} (\epsilon,\epsilon^\prime)
 \label{eq:chi4}
\end{equation}
where $\left( \chi^e_{sww} \right)^{-1}$ is the functional inverse of the density-density
correlation function defined as (see Eq. (3.5.8) of Ref.~\cite{Hansen1991}):
\begin{equation}
 \int_{-\infty}^{+\infty} d\epsilon^{\prime\prime} 
 \cdot
 \chi^e_{sww} (\epsilon,\epsilon^{\prime\prime})
 \left( \chi^e_{sww} \right)^{-1} (\epsilon^{\prime\prime},\epsilon^{\prime})
 =
 \delta ( \epsilon - \epsilon^\prime )
 \label{eq:chi5}
\end{equation}

With Eq.~\ref{eq:chi4} we can rewrite the HNC-like approximation of the indirect part of potential
of mean force (Eq.~\ref{eq:wHNC}) as:
\begin{equation}
 w^{e,HNC}_{sw}(\epsilon) 
 =
 -k_BT
 \left[
 \frac
 { 
  \rho^e_{sw}(\epsilon) - \rho^e_{sw,0}(\epsilon)
 }
 {
      \rho^e_{sw,0}(\epsilon) 
 }
 -
  \int_{-\infty}^{+\infty} d\epsilon^\prime 
    \cdot
    \left[ \rho^e_{sw}(\epsilon^\prime) - \rho^e_{sw,0}(\epsilon^\prime) \right]
    \cdot
    \left( \chi^e_{sww,0} \right)^{-1} (\epsilon,\epsilon^\prime)
  \right]
 \label{eq:wHNC3}
\end{equation}

\subsubsection{Percus-Yevick (PY) - like approximation.}

Again, following Percus \cite{Percus1962} we obtain the Percus-Yevick-like (PY-like) approximation
by
expanding the following functional:
\begin{equation}
 \rho^e_{sw}(\epsilon) e^{ \beta u^e_{sw}(\epsilon) } 
 \approx
 \rho^e_{sw,0}(\epsilon) 
 + 
 \int_{-\infty}^{+\infty} d\epsilon^\prime \cdot 
 \left( 
  \rho^e_{sw}(\epsilon^\prime) - \rho^e_{sw,0}(\epsilon^\prime)
 \right)
 \cdot
 \left.
 \frac
      {\delta \left[ \rho^e_{sw}(\epsilon) e^{ \beta u^e_{sw}(\epsilon) }
              \right] 
      }
      {\delta \rho^e_{sw}(\epsilon^\prime) }
  \right|_{\rho^e_{sw} = \rho^e_{sw,0}}      
 \label{eq:PY}
\end{equation}

We rewrite Eq.~\ref{eq:PY} via IPMF (Eq.~\ref{eq:we}):
\begin{equation}
 w^{e,PY}_{sw}(\epsilon) 
 =
 -k_BT 
 \ln
 \left(
 1
 +
  \int_{-\infty}^{+\infty} d\epsilon^\prime \cdot 
  \left[
 \frac
      {\delta (\epsilon - \epsilon^\prime) }
      {\rho^e_{sw,0}(\epsilon) }
    +
    \beta
    \left.
     \frac
      {\delta u^e_{sw}(\epsilon)  }
      {\delta \rho^e_{sw}(\epsilon^\prime) }
    \right|_{\rho^e_{sw} = \rho^e_{sw,0}}    
  \right]
 \right)
 \label{eq:wPY}
\end{equation}

With the help of Eq.~\ref{eq:wHNC} we can rewrite the PY-like approximation via the HNC-like
$w$:
\begin{equation}
w^{e,PY}_{sw}(\epsilon) 
=
 -k_BT 
 \ln
 \left(
 1
 -
 \beta w^{e,HNC}_{sw}(\epsilon) 
 \right)
 \label{eq:wPY2}
\end{equation}

\subsubsection{Lambda-integral in HNC-like approximation.}

When $u^e_\lambda$ is the solute-solvent interaction potential the corresponding IPMF is written as:
\begin{equation}
 w^{e,HNC}_{sw,\lambda}(\epsilon) 
 =
 -k_BT
 \left[
 \frac
 { 
  \rho^e_{sw,\lambda}(\epsilon) - \rho^e_{sw,0}(\epsilon)
 }
 {
      \rho^e_{sw,0}(\epsilon) 
 }
 -
  \int_{-\infty}^{+\infty} d\epsilon^\prime 
    \cdot
    \left[ \rho^e_{sw,\lambda}(\epsilon^\prime) - \rho^e_{sw,0}(\epsilon^\prime) \right]
    \cdot
    \left( \chi^e_{sww,0} \right)^{-1} (\epsilon,\epsilon^\prime)
  \right]
 \label{eq:wHNCl}
\end{equation}

With the linear dependence of $\rho^e_{sw,\lambda}$ on $\lambda$ (Eq.~\ref{eq:rhoel})
Eq.~\ref{eq:wHNCl} can be written via $w^{e,HNC}_{sw}$ at full solute coupling:
\begin{equation}
 w^{e,HNC}_{sw,\lambda}(\epsilon) 
 =
 \lambda 
 \cdot
 w^{e,HNC}_{sw}(\epsilon) 
 \label{eq:wHNCl2}
\end{equation}

The $\lambda$-integral in Eq.~\ref{eq:F3_2} can be written in HNC-like approximation as:
\begin{equation}
 \beta \int_{0}^1 d\lambda w^{e,HNC}_{sw,\lambda}(\epsilon) 
 =
 w^{e,HNC}_{sw}(\epsilon) \cdot \beta \int_{0}^1 d\lambda \cdot \lambda 
 =
 \frac{1}{2} \beta w^{e,HNC}_{sw}(\epsilon)
 \label{eq:intHNC}
\end{equation}

\subsubsection{Lambda-integral in PY-like approximation.}

When $u^e_\lambda$ is the solute-solvent interaction potential the corresponding IPMF in PY-like
approximation is written as (see Eq.~\ref{eq:wPY2}):
\begin{equation}
 w^{e,PY}_{sw,\lambda}(\epsilon) 
=
 -k_BT 
 \ln
 \left(
 1
 -
 \beta w^{e,HNC}_{sw,\lambda}(\epsilon) 
 \right)
=
 -k_BT 
 \ln
 \left(
 1
 -
 \lambda \cdot \beta w^{e,HNC}_{sw}(\epsilon) 
 \right) 
 \label{eq:wPYl}
\end{equation}

Now, we use the following known tabulated relation:
$$
\int dx \cdot \ln ( ax + b ) 
=
\frac{
       (ax+b) \cdot \ln ( ax + b ) - ax
     }
     {
       a
     }
$$

to find the $\lambda$-integral:
\begin{equation}
 \beta \int_{0}^1 d\lambda w^{e,PY}_{sw,\lambda}(\epsilon) 
 =
 -
 \frac{
         \left[ -\beta w^{e,HNC}_{sw}(\epsilon) + 1 \right] 
         \cdot
         \ln \left[ -\beta w^{e,HNC}_{sw}(\epsilon) +1 \right]
         +
         \beta w^{e,HNC}_{sw}(\epsilon)
      }
      {
         - \beta w^{e,HNC}_{sw}(\epsilon)
      }
 \label{eq:intPY}
\end{equation}

Next, we use the relation between $w$ in PY and HNC-like approximations at full solute coupling
$\lambda=1$ (see Eq.~\ref{eq:wPYl}):

\begin{equation}
 w^{e,PY}_{sw}(\epsilon) 
=
 -k_BT 
 \ln
 \left(
 1
 -
 \beta w^{e,HNC}_{sw}(\epsilon) 
 \right)
=> 
 -\beta w^{e,HNC}_{sw}(\epsilon) 
=
 e^{-\beta w^{e,PY}_{sw}(\epsilon)} - 1
 \label{eq:wPY1}
\end{equation}

Using Eq.~\ref{eq:wPY1} we rewrite Eq.~\ref{eq:intPY} as:
\begin{equation}
 \beta \int_{0}^1 d\lambda w^{e,PY}_{sw,\lambda}(\epsilon) 
=
 - \ln \left[ 1 - \beta w^{e,PY}_{sw}(\epsilon) \right] 
 + 1
 +
 \frac{    
         \ln \left[ 1 - \beta w^{e,PY}_{sw}(\epsilon) \right]  
      }
      {
         \beta w^{e,PY}_{sw}(\epsilon)
      }
 \label{eq:intPY2}
\end{equation}

\subsubsection{Constructing hybrid functional.}

The approximate functional is developed by Matubayasi and Nakahara \cite{Matubayasi2002} based on
the following considerations. To make an end-point expression of $\mu_{ex}$ we
need to approximate the $\lambda$-integral in Eq.~\ref{eq:F3}. Beforehand, we would like to note
that the approximate expression of the $\lambda$-integral combines four parts. 

Firstly, the
$\lambda$-integration can be analytically performed both in PY-like and HNC-like approximations (see
Eqs.~\ref{eq:intPY2} and \ref{eq:intHNC}). There is a general knowledge in the field that in the
case of simple liquids the PY approximation works better for short range repulsive potentials, while
HNC approximation performs better for long-range attractive potentials \cite{Hansen1991}. Matubayasi
and Nakahara \cite{Matubayasi2002} decided to use the PY-like expression for the $\lambda$-integral
in the unfavorable region of solvation ($w^e_{sw} \geq 0$) and HNC-like expression for the
$\lambda$-integral in the favorable region of solvation ($w^e_{sw} < 0$).

Secondly, the indirect part of potential of mean force $w^e_{sw}$ can be determined from unbiased
molecular simulations (regular MD or Monte-Carlo) only outside of the solute-core region (region
of very large solute-solvent interaction energies: $\epsilon$). Matubayasi and Nakahara
\cite{Matubayasi????} proposed to use HNC-approximation of the potential of mean force when
$w^e_{sw}$ is not resolved from molecular simulations. In HNC-like approximation $w^{e,HNC}_{sw}$
is determined by
the solute-solvent density distribution $\rho^e_{sw,0}$ and the inverse of the density-density
correlation function $\left( \chi^e_{sww,0} \right)^{-1}$ in the case of the zero solute-solvent
coupling (this can bee seen from Eq.~\ref{eq:wHNC3}, where the difference
$\rho^e_{sw} - \rho^e_{sw,0}$ can be safely approximated by $-\rho^e_{sw,0}$ since in the core
region $\rho^e_{sw} << \rho^e_{sw,0}$). Therefore, $w^{e,HNC}_{sw}$ in the
core-region can be calculated with high resolution in the
ensemble, where solute and solvent are fully decoupled and the
probability to find solvent molecule in the core region is high. The later can be in the most
convenient way realized by the insertion of the solute molecule configurations into the ensemble of
precalculated pure solvent configurations. 

Combination of the two different expression for the $\lambda$-integral and the two different input
$w$
functions results into functional consisting of four parts. 
The final expression for the excess chemical potential (Eq.~\ref{eq:Kcf_e_3}) is:

\begin{equation}
\mu_{ex} 
[ \rho^e_{sw}(\epsilon),\rho^e_{sw,0}(\epsilon),\chi^e_{sww,0}(\epsilon,\epsilon^\prime) ]
= 
\int_{-\infty}^{+\infty} d\epsilon
\rho^e_{sw} (\epsilon) \epsilon
- 
\mathcal{F} 
[ \rho^e_{sw}(\epsilon),\rho^e_{sw,0}(\epsilon),\chi^e_{sww,0}(\epsilon,\epsilon^\prime) ]
\label{eq:Kcf_e_4}
\end{equation}

where

$$
\mathcal{F} 
[ \rho^e_{sw},\rho^e_{sw,0},\chi^e_{sww,0} ]
= 
$$

\begin{equation}
=
k_BT 
\int_{-\infty}^{+\infty} d\epsilon
\left[
   \left(
	\rho^e_{sw} (\epsilon)- \rho^e_{sw,0}(\epsilon) 
    \right)
  - \rho^e_{sw}(\epsilon) \ln \frac{\rho^e_{sw}(\epsilon)}{\rho^e_{sw,0}(\epsilon)}
-
(\rho^e_{sw}(\epsilon) - \rho^e_{sw,0}(\epsilon))
\cdot
\mathcal{I}
[ \rho^e_{sw},\rho^e_{sw,0},\chi^e_{sww,0} ]
\right]
\label{eq:F3_2}
\end{equation}

where $\mathcal{I}$ is the approximated $\lambda$-integral:
\begin{equation}
\beta
\int_0^1 d\lambda
    w^e_{sw,\lambda} (\epsilon)
\approx
\mathcal{I}
[ \rho^e_{sw},\rho^e_{sw,0},\chi^e_{sww,0} ]
=
 \alpha(\epsilon) \cdot F_{w} (\epsilon)  
 + \left[ 1-\alpha(\epsilon) \right]
   \cdot F_{wHNC}(\epsilon) 
\end{equation}

where the functions $F_{w}$ and $F_{w^{HNC}}$ are in turn written as the combination of PY and
HNC-like expressions for the $\lambda$-integral:
\begin{equation}
    F_{w}(\epsilon) = 
    \begin{cases}
      \dfrac{\beta w^e_{sw}(\epsilon)}{2},& \text{when }  w^e_{sw}(\epsilon) \geq 0\\
      \beta w^e_{sw}(\epsilon) +1+\dfrac{\beta w^e_{sw}(\epsilon)}{e^{-\beta w^e_{sw}(\epsilon)}
-1},           & \text{when } w^e_{sw}(\epsilon) < 0
    \end{cases}
\end{equation}

and
\begin{equation}
    F_{wHNC}(\epsilon) = 
    \begin{cases}
      \dfrac{\beta w^{e,HNC}_{sw}(\epsilon)}{2},& \text{when }  w^{e,HNC}_{sw}(\epsilon) \geq 0\\
      -\ln \left[ 1 - \beta w^{e,HNC}_{sw}(\epsilon) \right] 
      + 1
      + \dfrac{\ln \left[ 1 - \beta w^{e,HNC}_{sw}(\epsilon) \right]}{\beta
w^{e,HNC}_{sw}(\epsilon)}
      ,
         & \text{when } w^{e,HNC}_{sw}(\epsilon) < 0
    \end{cases}
\end{equation}

The parameter $\alpha(\epsilon)$, which is responsible for merging parts with different $w$
functions, is set heuristically \cite{Matubayasi????} as \cite{Matubayasi2002}:
\begin{equation}
    \alpha(\epsilon) = 
    \begin{cases}
     1 ,& \text{when } \rho^e_{sw}(\epsilon) \geq \rho^e_{sw,0}(\epsilon) \\
     1-\left( \dfrac{\rho^e_{sw}(\epsilon) - \rho^e_{sw,0}(\epsilon)}{\rho^e_{sw}(\epsilon) +
\rho^e_{sw,0}(\epsilon)}  \right)^2  ,& \text{when } \rho^e_{sw}(\epsilon) < \rho^e_{sw,0}(\epsilon)
    \end{cases}
\end{equation}



\newpage
\bibliographystyle{unsrt}
\bibliography{frolov}

\end{document}